# Broadband Circular Polarizers Constructed by Helix-like Chiral Metamaterials

Ruonan Ji,[a,b,c] Shao-Wei Wang,[a,b, *] Xingxing Liu,[a,b,c] Xiaoshuang Chen[a,b] and Wei Lu[a,b]

[a] National Laboratory for Infrared Physics, Shanghai Institute of Technical Physics, Chinese Academy of Sciences, Shanghai 200083, China
[b] Shanghai engineering research center of energy-saving coatings, Shanghai 200083, China
[c] University of Chinese Academy of Sciences, Beijing 100049, China
* Corresponding author: wangshw@mail.sitp.ac.cn

In this paper, a kind of helix-like chiral metamaterial, which can be realized with multiple conventional lithography or electron beam lithographic techniques, is proposed to achieve broadband bianisotropic optical response analogous to helical metamaterial. On the basis of twisted metamaterial, via tailoring the relative orientation within the lattice, the anisotropy of arc is converted into magneto-electric coupling of closely spaced arc pairs, which leads to a broad bianisotropic optical response. By connecting the adjacent upper and lower arcs, the coupling of metasurface pairs is transformed to the coupling of the three-dimensional inclusions, and provides a much broader and higher bianisotropic optical response. For only a four-layer helix-like metamaterial, the maximum extinction ratio can reach 19.7. The operation band is in the wavelength range from 4.69 μm to 8.98 μm with an average extinction ratio of 6.9. And the transmittance for selective polarization is above 0.8 in the entire operation band. Such a structure is promising for integratable and scalable broadband circular polarizers, especially has great potential to act as broadband circular micropolarizers in the field of the full-stokes division of focal plane polarimeters.

## Introduction

Polarization imaging, especially circular polarization imaging is widely used in various fields, such as biomedical imaging, material sciences, space remote sensing, and military target recognition.[1] With the development of polarimeter, full-stokes division of focal plane (DoFP) polarimeters are regards as the advanced polarimeters. Full-stokes DoFP polarimeters utilize compact micropolarizer arrays integrated with the detector arrays to obtain linear and circular polarization information of different directions by only one measurement.[2, 3] Therefore, moving targets, such as flying airplanes and living cells can be detected using these polarimeters as imaging sensor.

However, nowadays, the micropolarizers of most DoFP polarimeters only consist of arrays of 0°, 45°, 90°, and 135° linear polarizers, which only capable of measuring the linear components of the Stokes vector sequentially.[4, 5] And the few available full-stokes DoFP polarimeters reports are using liquid crystal polymers as circular micropolarizers.[2, 3, 6] Either absorption based[7, 8] or interference-based[2], the LCP micropolarizer usually work at a specific wavelength in the visible or near-infrared range. In other words, liquid crystal polymers cannot be used as broadband circular micropolarizers or work in the infrared or terahertz, which are important operation bands for various applications of circular polarization imaging. In 2009, J. K. Gansel et.al proposed to prepare three-dimensional gold helices via direct laser writing followed by electrochemical deposition. An extinction ratio of about 10 in the wavelength range from 3.5-6.5 μm is realized.[9] Till now, circular polarizers based on helix structures are still believed as the broadest circular polarizer's candidates.[10] Various special fabrication process are proposed besides direct laser writing, such as focused ion beam induced-deposition (FIBID),[11] on-edge lithography (OEL),[12] glancing angle deposition (GLAD),[13] tomographic rotatory growth (TRG)[10] and so on. However, none of the above fabrication methods is compatible with other linear polarizer units' fabrication process to integrate on a chip, which is usually electron beam lithography or conventional lithographic techniques. Thus, the integration fabrication processes is an urgent problem to be solved for circular micropolarizers based on helix structures.

Metamaterials based on planer structures are more attractive candidates for many exciting applications as their much easier and mature fabrication processes.[14-16] In addition, they are usually realized with multiple conventional lithography or electron beam lithographic techniques, which are important fabrication processes for sub-wavelength gratings.[17-19] Thus, planer metamaterials are more promising candidates as circular micropolarizers. Spiral plasmonic nanoantennas are typical single layer planer metamaterials, they show narrow band bianisotropic optical response as the effect is mainly originated from the metal-insulator-metal (MIM) waveguide effect for the right-handed polarizations (RCP) and plasmonic vortex for the left-handed polarizations (LCP). According to the report of Bachman et.al., an extinction ratio of 5.1 is obtained at the main peak around 700 nm with a 4.2π model spiral array .[20] While for the bilayer twisted materials, such as conjugate gammadion,[21] twisted stakes,[22] twisted arcs,[23] twisted stacked L-shapes,[24] twisted split ring resonators (SRRs),[15, 16] twisted Fermat's spirals,[25] their selective effects to circular polarizations are mainly resulted from the coupling of upper and lower layers, which are still narrow band effect. In 2012, Zhao et.al reported a kind of twisted metamaterials based on closely spaced twisted broadband bianisotropic gold nanorod arrays. The operation band is successfully broadened to about 500nm (700-1200 nm), while the obtained maximum extinction ratio is only about 2.5.[26] Thus, the extinction ratio and operation band of planer metamaterials are very limited and far away from practical applications.

In this paper, an alternative venue to achieve broadband circular polarizers is proposed. Our design is based on a kind of helix-like planer chiral metamaterials, which not only has broadband bianisotropic optical response analogous to the helical metamaterials, but also reduces the fabrication difficulty. It largely enhances the practical probability for real-time Full-stokes polarization detection.

## Design and simulation method

Firstly, a metamaterial based on twisted nano arcs is considered. In each unit cell, the nano arcs are identical with same radius of $r$, width of $w$, thickness of $t$ and groove angle of $\vartheta$. And each arc shows strong surface anisotropy around the nanorod resonance. The arcs are cascading with a sub-wavelength separation distance $d$ and a specific rotation angle $\vartheta$ of the second surface compared with the first one. By closely staking the plasmonic metasurfaces and tailoring the relative orientation within the lattice, the anisotropy of each metasurface is converted into magneto-electric coupling of closely spaced twisted metasurface pairs. Then, arcs with the same size parameters, but smaller groove angle of $\beta$ are employed to connect the adjacent upper and lower arcs. The coupling of metasurface pairs is transformed to the coupling of the three-dimensional inclusions as broadband bianisotropic helical metamaterials. The schematic diagram of a four-layer helix-like chiral metamaterial is shown in Fig. 1.

In this paper, the simulations were performed by Finite Difference Time Domain (FDTD) method with commercial software (Lumerical FDTD Solutions) and CST Microwave Studio. The permittivity values in the infrared ranges of gold given by Palik[27] were used in the simulations. The structure was excited by a linear plane wave propagating along the negative $z$ direction with the electric field polarized in the $x(y)$ direction. Perfectly match layer (PML) absorbing boundaries were applied in the $z$ direction and periodic boundaries were used for a unit cell in the $x$-$y$ plane with a period of $p$. The transmission matrix for circularly polarized waves was calculated with the equation (1).

$$T_c = \begin{pmatrix} T_{rr} & T_{rl} \\ T_{lr} & T_{ll} \end{pmatrix} = \frac{1}{2}\begin{pmatrix} \left(T_{xx}+T_{yy}\right)+i\left(T_{xy}-T_{yx}\right) & \left(T_{xx}-T_{yy}\right)-i\left(T_{xy}+T_{yx}\right) \\ \left(T_{xx}-T_{yy}\right)+i\left(T_{xy}+T_{yx}\right) & \left(T_{xx}+T_{yy}\right)-i\left(T_{xy}-T_{yx}\right) \end{pmatrix} \quad (1)$$

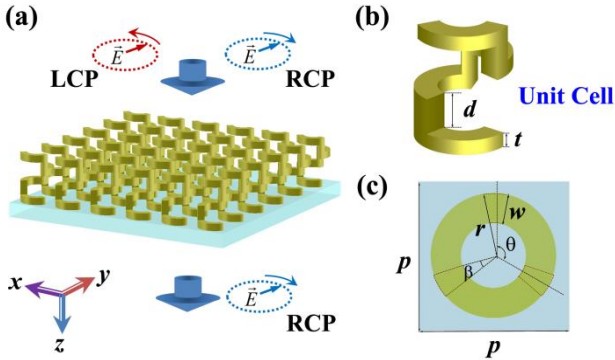

**Fig. 1** (a) the schematic diagram of the designed metamaterial selectively transmitted right-handed polarized incident waves，while left-handed polarized incident waves are almost completely blocked. (b) 3-D view of a unit cell of a four-layered helix-like chiral matematerials. (c) Top view of the unit cell.

Where the former and latter subscripts corresponding to the polarization of the transmittance and incident wave, and subscripts $r$ and $l$ represent the right-handed polarized (RCP) and left-handed polarized (LCP) waves respectively.

## Results and discussion

As an ideal circular polarizer, the selected incident circular polarization should be totally transmitted, whereas the unselected one is completely blocked. Here, the ratio of the selected polarization intensity to the blocked polarization intensity is defined as the extinction ratio, and the operation band of the circular polarizer is defined as the wavelength region in which the extinction ratio is no less than $\sqrt{2}$.[26] In addition, if pure circularly polarized transmitted light is required, the polarization conversion should be neglectable small.

As illustrates in Fig. 2(a), for the four-layer twisted metamaterials, the average extinction ratio in 2-2.76μm is 4.3, and the maximum extinction ratio can only reach 6.1. While for the four-layer helix-like metamaterials with same geometric parameters, as shown in Fig.2 (b), the operation band (2.59-6.21μm) can extend to above an octave, and the maximum and average extinction ratio can achieve 20.4 and 9.3, respectively. In other words, a four-layer helix-like metamaterial, has an analogous performance to the helical metamaterial mentioned in the Gansel's report (extinction ratio of about 10 in the wavelength range of 3.5-6.5μm)[9]. As for the polarization conversion shown in Fig. 2(c) and Fig. 2(d), the circular conversion transmittances of helix-like metamaterials stay below a few percentages for the entire calculated wavelength range with a maximum value of 0.06, much lower than the value of twisted metamaterials (maximum value of 0.13).

To understand the mechanism of the broadband bianisotropic optical response of helix-like chiral metamaterial, we started from the simplest bilayer case. The transmittance curves and surface current distributions of corresponding resonances are simulated and illustrated in Fig. 3. For the twisted arcs, with magneto-electric coupling between the metasurface pairs, resonant dips at $\lambda_{RT}$=1.75μm and $\lambda_{RT}$=2.25μm are excited by RCP and LCP incident waves, respectively (shown in Fig. 3(a)). According to hybridization model for chiral plasmonic Born–Kuhn modes[28], the LCP excites a lower energy level bonding mode with the same current directions of the two layers, while RCP excites a higher energy level anti-bonding mode with the opposite current directions (see Fig.3(b)). As mentioned in Ref. 28, LCP incident light first impinges on the upper arc, then rotates counter-clockwise and anti-aligns with the lower arc. Therefore, due to the symmetry of the structure, LCP excites a bonding mode with the same current directions of the two layers. Similarly, RCP rotates clockwise and excited an anti-bonding mode (see Fig. S1(a) in SI). Namely, by tailoring the relative orientation of arcs, the anisotropy of arc is converted into magneto-electric coupling of closely spaced arc pairs, which leads to a bianisotropic optical response. Though the dips are deep, the average extinction ratio and operation band are dramatically restricted as the resonance dips belonging to LCP and RCP are overlapped obviously.

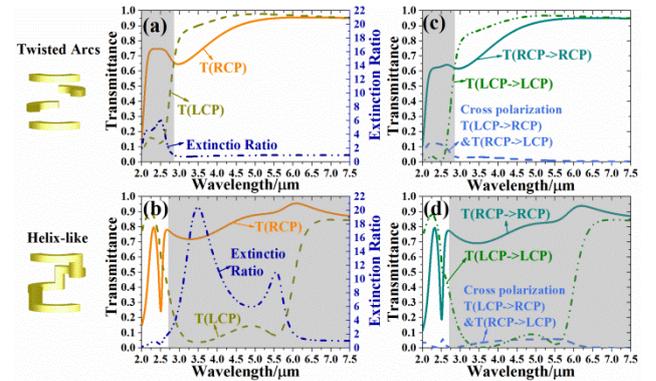

**Fig. 2** Transmittance and extinction ratio curves of a four-layered twisted arcs (a) and helix-like metamaterial (b). Corresponding polarization conversion spectra of twisted arcs (c) and helix-like metamaterial (d). The two structures are both left-handed and with same geometric parameters as follows, $t$=200 nm, $r$=500nm, $w$=200 nm, $d$=280 nm, $\vartheta$=120°, $\beta$=20° and $p$=1200 nm.

While for the helix-like metamaterial, an obviously resonant dip at $\lambda_{RH1}$=1.94μm and a slight dip at $\lambda_{RH2}$=3.91μm are excited by RCP incident waves, while two resonant dips at $\lambda_{LH1}$=1.73 μm and $\lambda_{LH2}$=3.59μm are excited by LCP ones. Similar with twisted arcs, the two sharper resonant dips at $\lambda_{RH1}$(1.94μm) and $\lambda_{LH2}$(3.59μm) are originated from anti-bonding(the current node is at the middle of the structure) and bonding mode (current direction is same in the whole structure), respectively (as shown in Fig. 3(d)). While the resonant dips at $\lambda_{LH1}$(1.73μm) and $\lambda_{RH2}$(3.91μm) are contributed by more complex and weaker resonant mode(see Fig. S1(b) in supplementary information). So, the resonant modes are corresponding to both arc pairs and connected arc, which shows a characteristic of three-dimensional inclusions. While with a left- handedness structure, the resonant dips excited by LCP incident waves is more obvious and deeper. Moreover, the effective length of the polarization charge oscillating of bonding mode is greatly increased after connecting, while the increase for anti-bonding mode is quite limited. As shown in Fig. 3(c), the resonant dip at $\lambda_{RT2}$=2.25μm in twisted arcs is red-shift to $\lambda_{LH2}$=3.59μm in helix-like metamaterial under LCP incident waves, while the one at $\lambda_{RT1}$=1.75μm is only red-shift to $\lambda_{RH1}$=1.94μm under RCP incident waves. Thus, in the helix-like metamaterials, the effective length of the polarization charge oscillating is tailored to separate the resonant dips belonging to bonding mode and anti-bonding mode. On the one hand, the extinction ratio is enhanced as the dips belonging to LCP and RCP are separated. On the other hand, the operation band is promising to be broadened as the resonant dips excited by LCP are separated as well.

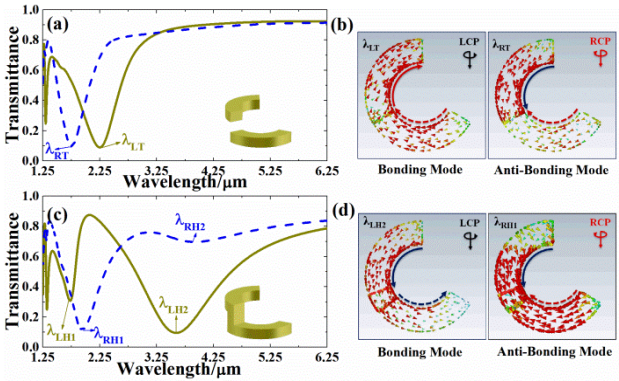

**Fig. 3** Transmittance curves under normal incident LCP and RCP waves of bi-layer twisted arcs (a) and helix-like (c) metamaterial. Surface currents induced in the structure by the corresponding polarized incident wave at resonant wavelengths marked in Fig. 3(a) of the bi-layer twisted arcs (a) and helix-like (b) metamaterials. The geometric parameters are same as Fig.2.

Then, the evolvement of transmittance curves under LCP and RCP incident waves with the increase of separation distance $d$ is studied. In twisted arcs, the magneto-electric coupling, which generates bianisotropic optical response, is existed in closely spaced arc pair. With the increase of separation distance, the coupling effect is attenuated. Thus, as shown in Fig. 4(a), the two resonant dips respectively excited by LCP and RCP are gradually overlapping with each other with the increase of separation distance $d$. When d>600nm, the bianisotropic optical response vanishes. Moreover, one may notice that only one resonant dip occurs when excited by LCP/RCP incident waves, as no extra metasurface exists between the two layers. As for the helix-like metamaterial, with a connected arc between the upper and lower arcs, more resonant dips occurs when the separation distance $d$ increasing. In addition, the bianisotropic optical response still exists even when the separation distance $d$ reaches 1.3μm. Thus, the magneto-electric coupling in helix-like metamaterial is a longer distance effect than that in twisted arcs. In the other words, in multi-layer helix-like metamaterial, the magneto-electric coupling not only exists in closely spaced arc pairs, but also the nonadjacent pairs. So, more complex resonant modes with energy level between pure bonding mode (all the current directions in the arcs are same) and pure anti-bonding mode (the current direction of a certain arc is opposite with its adjacent upper and lower arcs) will be excited.

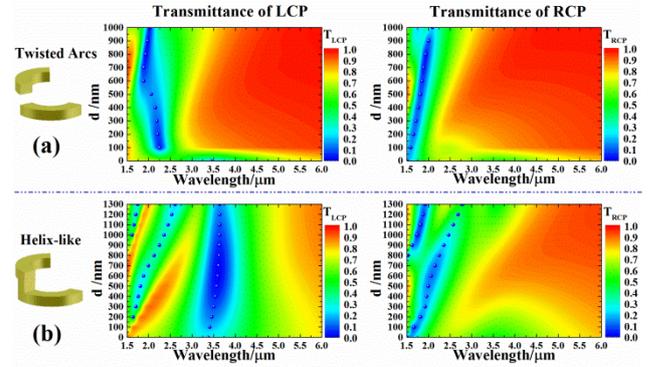

**Fig. 4** Transmittance of bilayer twisted arcs (a) and helix-like (b) metamaterials as functions of separation distance $d$. Structure parameters are same as Fig. 2, but the separation distance $d$ varies from 100 nm to 1000 nm.

How the bandwidth evolves with the increase of layer number is investigated. As mentioned above, in multi-layer helix-like metamaterial, more complex resonant modes with energy level between pure bonding mode and anti-bonding mode will be excited. And the overlapping of resonant dips will form a continuous low-transmittance band for LCP incident waves. Moreover, by tailoring the twisted angle and effective length of the polarization charge oscillating, one can guarantee that bonding mode is only excited by LCP incident waves. Thus, in multi-layer helix-like metamaterial, the longer edge of the operation band is determined by the resonant dip originated from bonding mode excited by LCP incident waves.

With the increase of layer numbers, the longer edge of the operation band is obviously shifting to longer wavelength as the effective length of the polarization charge oscillating of bonding mode is greatly increased, and the operation band is broadened as well. In addition, as shown in Fig. 5, the transmittance of LCP is greatly reduced with more layers staked. The minimum transmittance of LCP in a bi-layer helix-like metamaterial is 0.09, and the value is reduced to only 0.005 in a seven-layered case. And the maximum extinction ratio is increasing from 8.2 to 172.8 as well. However, one should notice that the optical losses of RCP incident waves also increase when excessive layers are staked. So, the layer number should be a compromise of extinction ratio, operation band and transmittance of RCP.

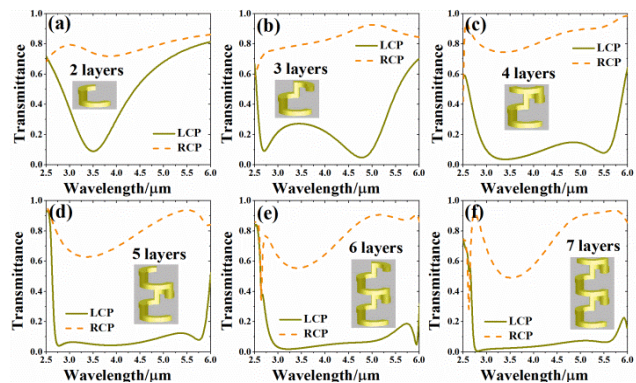

**Fig. 5** Normal incident transmission of helix-like metamaterials with different layer number of under LCP and RCP incident waves. Structure parameters are same as Fig. 2, but the layer numbers, *N*, is varied as indicated.

For simplicity, vacuum was applied as the spacers and no substrate was used in the above simulations. While in real applications, dielectric spacers and substrates should be considered. Simulated results show that when dielectric spacers and substrate (half-space geometry) are added, the operation band is red-shifted and obviously broadened. While the bianisotropic optical response slightly reduces as the optimal match is destroyed with the refractive index change of substrate and spacers (see Fig. S2 in SI). Moreover, the transmittance curves of thick and thin substrate are quite similar when the absorbance is ignored, except for the number of interference peaks and dips (see Fig. S3 and Fig. S4 in SI). So, in this paper, the thickness of substrate is set as 20μm. The transmittance and extinction ratio curves for a re-run optimized four-layered helix-like metamaterial with calcium fluoride[29] substrate and spacer is shown in Fig. 6. The transmittance of RCP can be maintained above 0.8 in the operation band of 4.69-8.98μm. And the maximum and average extinction ratios are 19.7 and 6.9, respectively.

## Conclusions

In this paper, a kind of helix-like chiral metamaterials is proposed as an alternative avenue to achieve integratable broadband circular polarizers. The structure of helix-like metamaterial is on the basis of twisted metamaterial. By stacking the arc arrays with a tailored rotational twist, the anisotropy of arc is converted into magneto-electric coupling of closely spaced arc pairs, leading to a broad bianisotropic optical response. Then, the adjacent upper and lower arcs are connected to transform the coupling of metasurface pairs into the coupling of the three-dimensional inclusions, which can be effectively operated like helical structures with much broader and higher bianisotropic optical response than twisted metamaterials. The transmittance of RCP can maintain above 0.8 in the operation band of 4.69-8.98μm. And the maximum and average extinction ratios are 19.7 and 6.9, respectively.

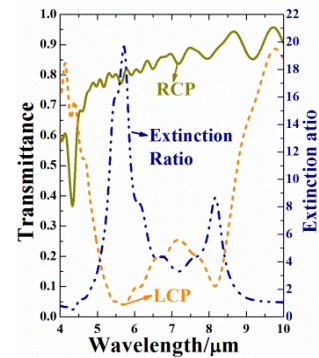

**Fig. 6** Normal incident transmittance curves and extinction ratio of helix-like metamaterials. Calcium fluoride is used as the substrate and spacer, and its permittivity index is given by Li etc[29]. Structure parameters are as follows, *t*=140 nm, *r*=450nm, *w*=120 nm, *d*=460 nm, *ϑ*=120°, *β*=20° and *p*=1200 nm. And the thickness of the substrate is $d_{sub}$=20μm.

Such a structure can be realized with multiple conventional lithography or electron beam lithographic techniques. It is promising to be applied in a variety of exciting applications for polarization control, especially has great potential to act as the integrated broadband circular polarizers for real-time full-stokes polarization detections.

## Acknowledgements

This work was partially supported by the Shanghai Science and Technology Foundations (13JC1405902, 15DZ2282100, 16DZ2290600), Youth Innovation Promotion Association CAS (2012189) and National Natural Science Foundation of China (61223006, 61376053).

# Supplementary Information

The hybridization model for chiral plasmonic arc pairs is shown in Fig. S1(a), LCP incident light first impinges on the upper arcs, then rotates counter-clockwise and anti-aligns with the lower arc. Therefore, due to the symmetry of the structure, LCP excites a bonding mode with same current directions of the two layers. Similarly, RCP rotates clockwise and excites an anti-bonding mode with a higher energy level.

The surface currents induced by the corresponding polarized incident waves at resonant wavelengths marked in Fig. 3(a) of the bi-layer helix-like metamaterials is shown in Fig.S1(b). At $\lambda_{RH1}$=1.94μm, the directions of induced surface currents are different at the upper part and lower part, and the current node occurs in the middle of the structure, which is quite similar with the anti-bonding mode in twisted arcs. And at $\lambda_{LH2}$=3.59μm, the directions of induced surface current are all same in the whole structure, corresponding to the bonding mode in the twisted arcs. While for the other two resonant dips at $\lambda_{LH1}$=1.73μm and $\lambda_{RH2}$=3.91μm, the resonant modes are more complex and with more than one current nodes.

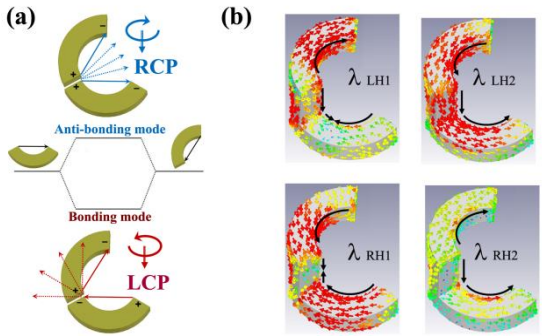

**Fig. S1** (a) Hybridization model for chiral plasmonic arcs pairs. (b) Induced Surface currents at resonant wavelengths marked in Fig. 3(a) of a bi-layer helix-like metamaterial. The geometric parameters as follows, $t$=200 nm, $r$=500nm, $w$=200 nm, $d$=280 nm, $\vartheta$=120°, $\theta$=20° and $p$=1200 nm.

Fig. S2 illustrates the influence of with and without substrate and spacers. In Fig.S2 (a), calcium fluoride (CaF$_2$) is used as the material of substrate (half-space geometry) and spacers. The permittivity index of calcium fluoride is given by Li etc[1]. After the dielectric spacers and substrate are added, the operation band obviously shifts to longer wavelength with a broader bandwidth. While as the former optimal match is destroyed as the effective refractive index changes, the bianisotropic optical response slightly reduces. The transmittance of LCP is increased while the transmittance of RCP is reduced slightly.

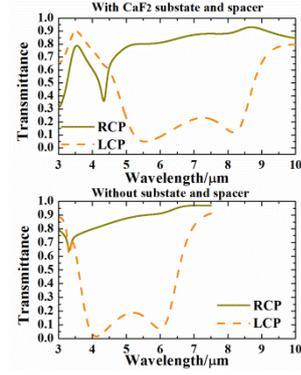

**Fig. S2** Transmittance curves of proposed four-layered helix-like metamaterial with calcium fluoride substrate and spacers(a) and without substrate and spacers(b). The geometric parameters are as Fig. S1.

The simulation results for transmittance curves of calcium fluoride substrate with different thicknesses are presented in Fig. S3. The results are obtained with commercial coating design software (CODE). As shown in Fig. S3, with the thickness increase of calcium fluoride, the transmittance curves are quite similar, except for the number increase of interference peaks.

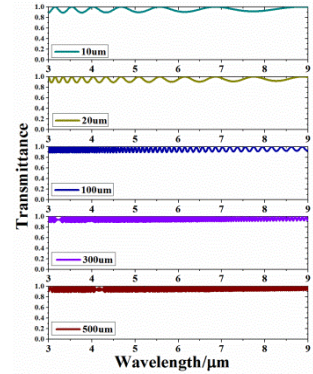

**Fig. S3** Transmittance curves of calcium fluoride substrate with different thickness.

Then, the transmittance and extinction ratio curves of proposed metamaterial with different substrate thicknesses are studied. As presented in Fig. S4, when the thickness of substrate is 0.2μm, no interference peaks are observed in the transmittance curves as the thickness of substrate is much smaller than the incident wavelength. The number of interference peaks is increased and the transmittance values fluctuate around the no interference case when the thickness of substrate increases. Similar phenomena are observed in the curves of extinction ratio. The thickness of substrate doesn't influence the overall performance of the device. There exists only a slightly fluctuation of extinction ratio at some certain wavelengths resulted from the interference.

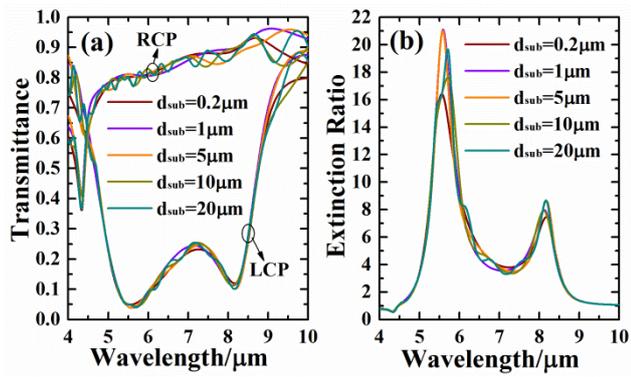

**Fig. S4** Transmittance and extinction ratio curves of proposed metamaterial with different thicknesses of substrate. Calcium fluoride is used as the material of substrate and spacers, and the thickness of the substrate is $d_{sub}$=20μm. Structure parameters are as follows, $t$=140 nm, $r$=450nm, $w$=120 nm, $d$=460 nm, $\vartheta$=120°, $\beta$=20° and $p$=1200 nm.